\documentclass[a4paper,11pt]{article}

\pdfoutput=1 % if your are submitting a pdflatex (i.e. if you have
\usepackage{jheppub} % for details on the use of the package, please
\usepackage[T1]{fontenc} % if needed

\usepackage{enumerate}
\usepackage{graphicx}% Include figure files
\usepackage{dcolumn}% Align table columns on decimal point
\usepackage{bm}% bold math
\usepackage{mathrsfs}
\usepackage{amsmath}
\usepackage[usenames,dvipsnames]{color} 
\usepackage{amssymb}
\usepackage{slashed}
\usepackage{diagbox}
\usepackage{subfigure}
\usepackage[table]{xcolor}
\usepackage{color,soul} 
\setul{0.5ex}{0.3ex}
\setulcolor{red}
\setstcolor{red}

\usepackage{subfigure} 
\usepackage{color,soul} 
\setul{0.5ex}{0.3ex}
\setulcolor{red}
\setstcolor{red}

\usepackage{tikz}
\usetikzlibrary{calc}

\usepackage{multirow}

\usepackage[normalem]{ulem}

\usepackage[mathlines]{lineno}% Enable numbering of text and display math

\title{\boldmath Cosmic structure formation in a flavoured U{$\left(1\right)$} dark sector at
small scales }

\author[a]{Ottavia Balducci,}
\author[a]{Stefan Hofmann,}
\author[a]{and Alexis Kassiteridis}

\affiliation[a]{Arnold Sommerfeld Center for Theoretical Physics,\\ Theresienstra{\ss}e 37, 80333 M\"unchen}

\emailAdd{ottavia.balducci@physik.uni-muenchen.de}
\emailAdd{stefan.hofmann@physik.uni-muenchen.de}
\emailAdd{a.kassiteridis@physik.uni-muenchen.de}

\abstract{ In this work we study the footprint of a secluded dark
  sector with a U{$\left(1\right)$} gauge symmetry on the formation of
  cosmic structure at small scales. A single generation of dark
  fermions with mass in the TeV range is able to elegantly solve all
  small-scale issues of structure formation, while still being
  consistent with the cosmological history of the universe. A new
  interaction at the MeV scale is introduced for this purpose. As a
  generalization we consider multiple generations of fermions as well
  and show that such extensions are also cosmologically viable and
  allow a more flexible choice of the parameters of the theory. }

\begin{document}
\maketitle
\flushbottom

\section{Introduction}
Almost all microscopic and macroscopic phenomena that we experience
arise directly from a single abelian interaction between electrical
charges, while non-abelian interactions become only accessible at a
considerably higher energy-scale.  Below this scale visible matter is
dominated by a U{$\left(1\right)$} gauge symmetry giving rise to
electromagnetism.  Inspired by its success, we investigate a secluded
sector with a U{$\left(1\right)$} symmetry in respect of cosmic
structure formation.  The secluded sector could be a dark completion
of the Standard Model or arise as an effective description of some
symmetry established at energies higher than those considered here and
suppressed accordingly.

There is a plethora of dark matter particle candidates in the
literature.  Recently, considerable attention has been devoted to dark
sectors that allow for a consistent history and hierarchy of cosmic
structure formation. In particular, the cosmological concordance
($\Lambda$CDM) model faces challenges known in the literature as cusp
versus core, missing satellites, too big to fail and diversity
problems, all related to structure formation on small scales, where
the particle nature of dark matter and its interactions are absolutely
relevant.  For a thorough discussion see
Refs. \cite{Aarssen:2012fx,Oman:2015xda}.  This work gives a simple
solution to these problems in accordance with constraints from
astroparticle physics.

For simplicity we consider the low-energy limit of a dark sector with
a spontaneously broken abelian symmetry. In addition we assume that
the abelian interaction scale is sufficiently lower than the energy
scales characterising all other forces in the dark sector such that
they can be neglected for the purposes of this article.  This results
in an effective dark sector similar to a secluded version of quantum
electrodynamics with the crucial difference of a massive force
mediator instead of a massless one, see Ref. \cite{Tulin:2017ara} for
its phenomenology.  In addition to the Proca mediator, its field
content includes $N$ heavier and $N$ lighter dark fermions which
transform as singlets under the gauge symmetries of the Standard
Model.  Provided that the dark matter mass lies in the TeV range, kinetic
decoupling of dark matter from its relativistic scattering partners is
scheduled for after the epoch of big bang nucleosynthesis, which is
essential for a consistent small-scale formation of cosmic structure
\citep{Aarssen:2012fx} within the cosmological concordance model.  We
show that already $N=1$ presents a solution to the above mentioned
problems.

Dark sectors under the spell of a new U$(1)$ symmetry have been
investigated in the literature \cite{Bringmann:2013vra, Ko:2014bka,
  ElHedri:2018cdm}. Let us comment on the differences to the dark
sector suggested in this work.  Previous models are not strictly
compatible with constraints from big bang nucleosynthesis and/or from
observations of the cosmic microwave background radiation, do not
contain flavoured dark matter, do not accommodate a gauge-invariant
mechanism for mass generation in their ultraviolet completion
\cite{Dasgupta:2013zpn}, or are less minimal compared to the dark
sector presented here in regard to the field content.  It is also
important to mention that while mirror dark matter \cite{Foot:2014mia}
also postulates a new U{$\left(1\right)$} symmetry for the dark matter
particles, it considers a massless mediator in the dark sector, which
leads to different cosmological
predictions. Furthermore, models departing from the CDM
  perspective, e.g. theories with feebly interacting DM
  \cite{Hager:2020une}, can lead to similar results as well.

This paper is organised as follows. In Section \ref{motivation} we begin with a brief discussion of our assumptions 
underlying the dark sector presented in 
Section \ref{lagrangian-cross-sections}, followed by a discussion of parameter constraints in 
Section \ref{constraints}. Finally, in Section
\ref{cosmological-observables} we discuss the cross sections for elastic
scattering and annihilation processes and calculate the cosmological
observables which are relevant for solving the small-scale
problems. We present our results and conclusions in Section
\ref{results}.

\section{Assumptions\label{motivation}}

We summarise the key assumptions which lead to the dark sector
described in the following section.
\begin{enumerate}[(1)]
\item {\bf Coupling:}
The dark sector was in local thermal equilibrium
with the Standard Model only well before the epoch of big bang
nucleosynthesis and is secluded since then.
\label{assum1}
\item {\bf Visible entropy:}
Only degrees of freedom carried by Standard Model fields contribute to the entropy of the visible sector 
after its decoupling from the dark sector.
\label{assum2}
\item {\bf Dark matter:}
The dark matter relic density is dominated by stable particles with masses in the TeV range. 
Mixed dark matter scenarios are not considered.
\label{assum3}
\item {\bf Dark abelian gauge symmetry:}
The dark sector is anomaly-free, but the abelian charges of the dark fermions are otherwise arbitrary. 
The abelian interaction scale is assumed to be in the MeV range.
\label{assum4}
\item {\bf Symmetry breaking:} The spontaneous symmetry breaking in
  the dark sector is assumed to happen well before the Standard Model
  electroweak phase transition.
  \label{assum5}
\end{enumerate}

Let us justify these assumptions. The lack of any detection evidence
hints at a dark sector which could have been in local thermal
equilibrium with the Standard Model sector only well before the epoch
of big bang nucleosynthesis and is now completely secluded, hence
assumption (\ref{assum1}).  While this is sometimes referred to as a
nightmare scenario, because no interactions with Standard Model
particles can ever be observed and hence no direct detection will be
technically feasible, such a dark sector can nevertheless be tested by
comparing its predictions for the abundance of cosmic structures and
their hierarchical formation history with observations.  Assumption
(\ref{assum2}) should be self-evident in the context given here.  It
is known that a dark matter particle candidate with a mass in the TeV
range can in principle give rise to a consistent hierarchical
formation of cosmic structure within the $\Lambda$CDM concordance
model, provided that at least one new interaction with an energy scale
well below the electroweak scale is operative in the dark sector
\cite{Tulin:2013teo,Tulin:2017ara}. The other requirements in
assumption (\ref{assum3}) are listed for simplicity (not for
necessity) and can be lifted for the price of a more technical
presentation.  The last requirement in assumption (\ref{assum4}) has
been justified above. The remaining statement concerns quantum
consistency and is a matter of taste. Finally, assumption
(\ref{assum5}) ensures that the dark fermions have already attained
their mass at the time of freeze-out and thus the correct relic
density is recovered.

\section{The dark sector}\label{lagrangian-cross-sections} 
We suggest the following particle spectrum for the dark sector:
$2N$ fermions $\chi^{\left(j\right)}$ and $\chi^{\prime\,{}\left(j\right)}$
with $j\in\{1,\ldots,N\}$ which transform as singlets under the gauge 
transformations of the Standard Model. 
We will show below that 
quantum consistency (assumption (\ref{assum4})) demands an 
even number of dark fermions. 
The dark spectrum contains a massive vector boson $X$ 
which prior to spontaneous symmetry breaking is the gauge field
associated with the new local abelian symmetry imposed on the dark sector. 
In order to distinguish it from the abelian gauge symmetry of the Standard Model 
(prior to spontaneous symmetry breaking) we denote it by U{$\left(1\right)^{\prime}$}.
In addition, the dark sector contains a complex scalar $\tilde{H}$ which mimics 
the Higgs boson $H$ in the Standard Model. 
Before elaborating further, let us define the dark sector via its Lagrange density:

\begin{eqnarray}
  \mathcal{L}&=&\sum_{j=1}^{N}\overline{\chi_{\rm{}L{}}^{\left(j\right)}}{\rm{i}}\slashed{D}\chi_{\rm{}L}^{\left(j\right)}+\sum_{j=1}^{N}\overline{\chi_{\rm{}R{}}^{\left(j\right)}}{\rm{i}}\slashed{D}\chi_{\rm{}R}^{\left(j\right)}+\sum_{j=1}^{N}\overline{\chi_{\rm{}L{}}^{\prime\,{}\left(j\right)}}{\rm{i}}\slashed{D}\chi_{\rm{}L}^{\prime\,{}\left(j\right)}+\sum_{j=1}^{N}\overline{\chi_{\rm{}R{}}^{\prime\,{}\left(j\right)}}{\rm{i}}\slashed{D}\chi_{\rm{}R}^{\prime\,{}\left(j\right)}\nonumber{}\\
  &&-\sum_{j=1}^{N}\overline{\chi_{L}^{\left(j\right)}}\frac{\tilde{H}}{v_{\rm{}d}}m_{\chi}^{\left(j\right)}\chi_{R}^{\left(j\right)}-\sum_{j=1}^{N}\overline{\chi_{L}^{\prime\,\left(j\right)}}\frac{\tilde{H}}{v_{\rm{}d}}m_{\chi^{\prime}}^{\left(j\right)}\chi_{R}^{\prime\,\left(j\right)}-{\rm{}H.c.}\label{ds}\\
  &&-\frac{1}{4}X^{\mu\nu}X_{\mu\nu}-\frac{\epsilon}{2}B^{\mu\nu}X_{\mu\nu}+\left(D_{\mu{}}\tilde{H}\right)^{\ast}D^{\mu}\tilde{H}-\frac{\lambda}{2}\left(\tilde{H}^{\ast}\tilde{H}-v_{\rm{}d}^{2}\right)^{2}+a^{2}\tilde{H}^{\ast}\tilde{H}H^{\ast}H\;{}.\nonumber{}
\end{eqnarray} 
Here, {$\chi^{\left(\prime\right)}_{L}$} and
{$\chi^{\left(\prime\right)}_{R}$} are the left- and right-chiral
projections of the fermionic fields, respectively. The
U{$\left(1\right)^{\prime}$}-covariant derivative is given by
{$D_{\mu}=\partial_{\mu}-{\rm{}i}X_{\mu}Q$}, where {$Q$} is the charge
operator corresponding to the {${\rm{}U}\left(1\right)^{\prime}$}
symmetry in the dark sector. The dark fermions have chiral charges
{$Q_{L}^{\left(j\right)}$}, {$Q_{R}^{\left(j\right)}$},
{$Q_{L}^{\prime\,\left(j\right)}$}, and
{$Q_{R}^{\prime\,\left(j\right)}$}, and {$\tilde{H}$} has charge
{$Q_{\tilde{h}}$} (the subscript $\tilde{h}$ refers to the expansion
of $\tilde{H}$ around its vacuum expectation value, as explained
below).  The dark fermion masses are denoted by
{$m_{\chi}^{\left(j\right)}$} and
{$m_{\chi^{\prime}}^{\left(j\right)}$}, respectively, and we assume,
without loss of generality, the following mass relations: For each
$i\in\{1,\ldots,N\}$ let
{$m^{\left(i\right)}_{\chi}>m_{\chi^{\prime}}^{\left(j\right)}$} for
all {$i,j\in\left\{1,\ldots{},N\right\}$}.  In other words, the
{$\chi$} fermions are the dark matter particles, while the
$\chi^{\prime}$ fermions are their relativistic scattering partners in
the dark sector. This designation is in accordance with assumption
(\ref{assum3}).  {$B^{\mu\nu}$} denotes the field strength tensor of
the vector boson {$B_{\mu}$} that corresponds to the abelian gauge
symmetry of the Standard Model before electroweak symmetry breaking,
and {$X_{\mu\nu}=2\partial_{\left[\mu\right.}X_{\left.\nu\right]}$} is
the field strength of the dark vector boson.  The dimensionless
parameter {$\epsilon$} is a priori arbitrary and describes the
strength of the kinetic mixing between the abelian force mediators in
the visible and the dark sector. Due to this kinetic-mixing
contribution all fermions in the dark sector receive electric charges
{$Q_{\rm{}L}\epsilon$} and {$Q_{\rm{}R}\epsilon$}.  Therefore
{$\epsilon$} must be strongly constrained a posteriori for
phenomenological reasons, as will be explained in Section
\ref{constraints}.  $v_\mathrm{d}$ denotes the vacuum expectation
value attained by $\tilde{H}$.  Finally, {$a$} is another free
dimensionless parameter describing the coupling strength of the
contact between the complex dark scalar $\tilde{H}$ and the Higgs
boson $H$ of the Standard Model.

Let us elaborate on the role of $\tilde{H}$. After the spontaneous
breaking of the U{$\left(1\right)^{\prime}$} symmetry {$\tilde{H}$}
can be decomposed in its vacuum expectation value $v_\mathrm{d}$ and
fluctuations around it,
{$\tilde{H}\left(x\right)=v_{\rm{}d}+\left(\tilde{h}\left(x\right)+i\tilde{h}^{\prime}\left(x\right)\right)/\sqrt{2}$},
where {$\tilde{h}$} and {$\tilde{h}^{\prime}$} are real scalar fields.
The dark fermions thus obtain the Dirac masses
{$m_{\chi}^{\left(i\right)}$} and
{$m_{\chi^{\prime}}^{\left(j\right)}$}, respectively, through the
Yukawa potential in (\ref{ds}). By naturalness the Yukawa couplings
must satisfy
{$m_{\chi^{\left(\prime\right)}}^{\left(i\right)}/v_{\rm{}d}\lesssim{}1$}
and this implies that {$v_{\rm{}d}$} must be at least in the TeV
range.  Furthermore, the kinetic term
$\left(D_{\mu}\tilde{H}\right)^{\ast}D^{\mu}\tilde{H}$ gives a
contribution {$Q_{\tilde{h}}^{2}v_{\rm{}d}^{2}X_{\mu}X^{\mu}$} after
symmetry breaking, corresponding to a mass
{$M_{X}=Q_{\tilde{h}}v_{\rm{}d}$} for the dark gauge boson.  Note that
Dirac fermions with different chiral charges for the left- and
right-handed particles are necessary in order to obtain a massive
vector boson, since the Yukawa potential in Eq. (\ref{ds}) determines
the following relation between the chiral charges and the charge of
{$\tilde{H}$},
\begin{eqnarray}
\pm Q_{\tilde{h}}&=&Q_{L}^{\left(j\right)}-Q_{R}^{\left(j\right)}\quad{}\, , \label{higgs-charge1}\\
\pm Q_{\tilde{h}}&=&Q_{L}^{\prime\,{}\left(j\right)}-Q_{R}^{\prime\,{}\left(j\right)}\label{higgs-charge2}\; ,
\end{eqnarray}
for all generations of dark fermions.  Since {$Q^{\left(j\right)}$}
and {$Q^{\prime\,\left(j\right)}$} must be sufficiently large to yield
the correct relic density for dark matter (see below), and $M_X$ is
required to be in the MeV range by assumption (\ref{assum4}),
{$Q_{\tilde{h}}$} must be much smaller than {$Q^{\left(j\right)}$} and
{$Q^{\prime\,\left(j\right)}$}.  After spontaneous symmetry breaking
in the dark sector, the quartic potential for {$\tilde{H}$} contains a
term {$-\lambda{}v_{\rm{}d}\tilde{h}^{2}$}, which corresponds to a
mass {$M_{\tilde{h}}=\sqrt{2\lambda}v_{\rm{}d}$} for $\tilde{h}$ after
the phase transition. Since the temperature of the symmetry breaking
is proportional to the mass of the dark Higgs field up to factors of
order one and since {$v_{\rm{}d}\sim{}$} TeV, {$\lambda{}\sim{}1$}
must hold by assumption (\ref{assum5}).

By assumption (\ref{assum4}), the dark sector is anomaly-free, implying the following 
conditions:
\begin{eqnarray}
\sum_{j=1}^{N}(Q_{L}^{\left(j\right)}-Q_{R}^{\left(j\right)})&=&\sum_{j=1}^{N}(Q_{R}^{\prime\,{}\left(j\right)}-Q_{L}^{\prime\,{}\left(j\right)})\label{anomaly1}\,{},\\
\sum_{j=1}^{N}(Q_{L}^{\left(j\right)\,{}3}-Q_{R}^{\left(j\right)\,{}3})&=&
\sum_{j=1}^{N}(Q_{R}^{\prime\,{}\left(j\right)\,{}3}-Q_{L}^{\prime\,{}\left(j\right)\,{}3})\,{}.
\label{anomaly2}
\end{eqnarray}
In principle we could have started with $N$ generations of heavy and $N^\prime$ generations 
of lighter dark fermions, in which case the summation on the left hand side of (\ref{anomaly1}) and (\ref{anomaly2})
would extend over $N$ generations while the summation on the right hand side would
extend over $N^\prime$ generations. 
Due to the charge relations (\ref{higgs-charge1}) and (\ref{higgs-charge2}), $N=N^\prime$ is enforced.
This explains why the total number of dark fermion generations had to be even.
There are then two solutions of the quantum consistency conditions (\ref{anomaly1}) and (\ref{anomaly2}):
\begin{eqnarray}
Q_{L}^{\left(j\right)}&=&Q_{R}^{\prime\,{}\left(j\right)}\;{},\;{}Q_{R}^{\left(j\right)}=Q_{L}^{\prime\,\left(j\right)} \quad \text{and}\\
Q_{L}^{\left(j\right)}&=&-Q_{L}^{\prime\,\left(j\right)}\;{},\;{}Q_{R}^{\left(i\right)}=-Q_{R}^{\prime\,\left(j\right)}\label{sol1}\;{}.
\end{eqnarray}
Without loss of generality we choose solution (\ref{sol1}).

Finally, only the Higgs portal and the kinetic mixing in (\ref{ds}) give rise to particle interactions 
between the Standard Model and the dark sector. However, the corresponding couplings 
are so severely constrained, as will be shown in the following sections, that the 
dark sector is effectively a secluded sector.

\section{Constraining the parameter set\label{constraints}}
In this section we present constraints on the parameters of the dark sector (\ref{ds}).
\begin{itemize}
	\item[1.] {\bf Dark matter mass:}
	Requiring partial-wave unitarity \cite{Griest:1989wd} leads to an upper bound 
	{$m_{\chi}^{\left(j\right)}<\mathcal{O}\left(10^2\right)$} TeV on the
	mass of the dark matter particles.
	\item[2.] {\bf Couplings:} The effectiveness of the
          interactions between the Standard Model and the dark sector
          depends strongly on the value for the bridge coupling {$a$}
          (as well as on the dark Higgs mass).  Bounds on the coupling
          {$a$} can be found in Ref.  \cite{Feng:2017vli}. Assumption
          (\ref{assum5}) implies that {$v_{\rm d}\gtrsim M_{\tilde{h}}
            \sim $} TeV must hold, where $M_h$ denotes the mass of the
          Higgs boson in the Standard Model. This implies an upper
          bound on the bridge coupling: {$a\lesssim{}10^{-2}$}.
	
	The
	strongest constraints on the coupling for kinetic mixing
	{$\epsilon{}$} come from observations of supernova SN1987A and require
	{$\epsilon{}\lesssim{}10^{-13}$} \cite{Kazanas:2014mca}.
	\item[3.] {\bf Dark charges:} Let us define $g_i
          :=\tfrac{1}{2}(Q_{L}^{\left(i\right)}+Q_{R}^{\left(i\right)})$
          and approximate
          {$g_{i}\approx{}Q_{L}^{\left(i\right)}\approx{}Q_{R}^{\left(i\right)}$},
          which is valid to lowest order in $Q_h \ll g_i$. Similarly,
          {$g_{i}^{\prime}:=
            \tfrac{1}{2}(Q_{L}^{\prime\,\left(i\right)}+Q_{R}^{\prime\,\left(i\right)})$},
          and
          {$g_{i}^{\prime}\approx{}Q_{L}^{\prime\,\left(i\right)}\approx{}Q_{R}^{\prime\,\left(i\right)}$}.
          Dark matter particles were thermally produced. The condition
          for the heavier dark matter fermions $\chi$ to be in local
          thermal equilibrium with their relativistic scattering
          partners $\chi^\prime$ (before annihilation processes
          commence effectively) becomes
	\begin{equation}
		4\pi \sum_{j=1}^N \alpha_{ij^{\prime}} := \sum_{j=1}^Ng_i g_j^\prime >
		\left(\frac{m^{\left(i\right)}_\chi{}x_{f}}{ M_{\rm Pl}}\right)^{\frac{1}{2}}\, ,\label{wimp-condition}
	\end{equation}
	where $x_{\rm f} := m_{\chi}^{\left(i\right)}/T_{\rm f}$ is the dimensionless chemical decoupling parameter 
	for the $i$th dark matter particle. 
	\item[4.] {\bf Effective degrees of freedom:}
	The Standard Model assigns $N_\nu = 3.046$ effective degrees of freedom to its neutrinos (collectively 
	denoted by $\nu$), but the presence of light fermions in the dark sector can lead to a deviation 
	$\Delta N_\mathrm{eff}$ from this number and therefore from the effective number of degrees of freedom 
	invested in all relativistic particles at a given cosmological epoch. 
	For the dark sector (\ref{ds}), the deviation is given by
	\begin{equation}\label{delta effective}
	\Delta N_{\rm eff}= N_\nu
	\frac{\rho_{\rm{}rel}^{\left({\rm{}DS}\right)}}{\rho_\nu}\;{},
	\end{equation}
	where $\rho_\nu$ and $\rho_{\rm{}rel}^{\left({\rm{}DS}\right)}$
	are the energy density of Standard Model neutrinos and the energy density 
	of the relativistic particles in the dark sector, respectively. 
	Entropy conservation gives 
	\begin{equation}
	\varepsilon \vert_{T_\nu} : = \left(\frac{T_{\rm DS}}{T_{\nu}}\right)^{3} = \frac{g^*_{\rm DS}(T_D)}{g^*_{\rm SM}(T_D)} \frac{g^*_{\rm SM}		(T_{\nu})}{g^*_{\rm DS}(T_\nu)}\, ,\label{epsilon}
	\end{equation}
	where $T_{\nu}$ is the temperature of the Standard Model neutrinos, $T_\mathrm{DS}$ denotes the temperature 
	of the relativistic fermions in the dark sector, $g^*_\mathrm{SM}$ and $g^* _{\rm DS}$ are the entropy 
	degrees of freedom of the Standard Model and of the dark sector, respectively, at the given temperatures,
	where $T_\mathrm{D}$ is the temperature when the visible and dark sectors decouple.  
	Note that assumption (\ref{assum2}) has been used to establish the equality in (\ref{epsilon}). 
	We take $T_{\nu\mathrm{D}}=2.3$MeV for the neutrino temperature at decoupling \cite{Enqvist:1991gx}.
	With the scaling relation (\ref{epsilon}) follows $\Delta N_{\rm eff}= N_\nu (\varepsilon \vert_{T_\nu})^{4/3}$.
	The upper bound {$a\lesssim{}10^{-2}$} on the bridge coupling implies that 
	the dark sector is no longer in local thermal equilibrium 
	at the electroweak phase transition.
	
	Table \ref{mass-spectra} shows the values of
	{$\left.\Delta{}N_{\rm{}eff}\right|_{\rm{}BBN}$} for {$N=1$} and
	{$N=2$}. For one generation of dark matter particles, 
	there is a unique mass hierarchy 
	$m_{\chi^\prime} < T_\mathrm{BBN} < T_\mathrm{D} < m_\chi$. 
	The first relation guarantees a sufficiently late kinetic decoupling of dark matter particles 
	from their relativistic scattering partner to ensure a consistent formation of dark matter
	structures on small scales. The last relation follows from assumption (\ref{assum3}). 
	The case of two generations allows for multiple mass
        hierarchies and all the ones in accordance with our
        assumptions are listed in Table \ref{mass-spectra}. The hierarchies II-IV 
	are excluded since the Planck collaboration \cite{Aghanim:2018eyx} gives the constraint
	{$\left.\Delta{}N_{\rm{}eff}\right|_{\rm{}BBN}\leq{}0.234{}$} for a
	deuterium fraction computed with \texttt{PRIMAT} \cite{Pitrou:2018cgg}.
	Let us stress that IV could still be allowed, if the more relaxed constraint
	{$\left.\Delta{}N_{\rm{}eff}\right|_{\rm{}BBN}\leq{}0.374$} \cite{Aghanim:2018eyx}
	is considered instead. The difference in these constraints can be traced back 
	to the question whether the helium fraction is allowed to vary independently from 
	the measured deuterium abundance as reported in \cite{Cooke:2017cwo}.
	The first two columns in Table \ref{mass-spectra} pass both constraints.
	
	Finally, there is a relation between {$\Delta{}N_{\rm eff}$} and the mass 
	{$m_{\chi^\prime}$} of the lightest dark fermion.
	Let us consider the case $N=1$ for simplicity with the mass hierarchy
	$m_{\chi^\prime} < T_\mathrm{BBN} < T_\mathrm{D} < m_\chi$
	as given in Table \ref{mass-spectra}. 
	The $\chi^\prime$ are the last relativistic scattering partners 
	of the dark matter particles, and their interaction determines
	the kinetic decoupling temperature $T_{\rm kd}$. 
	Since a later kinetic decoupling is beneficial for a sound formation of 
	dark matter structures on small scales, we may assume 
	that $\chi^{\prime}$ is still relativistic (or at least close to it)
	around kinetic decoupling, so $m_{\chi^{\prime}} \lesssim T_{\rm kd}$. 
	After chemical decoupling, the contribution of these particles to 
	the energy density of the Universe is $\Omega_{\chi^{\prime}} h^2 \approx \,
	m_{\chi^\prime}/{255\, \rm eV}$, assuming they are stable and due to 
	assumption (\ref{assum1}). As stated in (\ref{assum3}), we refrain here 
	from considering mixed dark matter scenarios.
	Therefore, {$\Omega_{\chi^\prime}h^{2}<0.0245$} \cite{Ade:2015xua}
	is a conservative bound corresponding to a minimum dark matter mixture 
	in the relic density. This translates into an upper bound 
	$m_{\chi^\prime}\lesssim 0.65 $ eV. Assuming instantaneous recombination 
	at $T=0.3$eV, the contribution of the lightest dark fermion to the 
	energy budget gives
	\begin{eqnarray}
	\left.\frac{\Delta N_{\rm eff}}{\varepsilon^{4/3}}\right\vert_\mathrm{rec}
	=
	\frac{240}{7\pi^4}
	\int\limits_{x_{{\rm rec}}}^\infty
	{\rm d}z\; z^2 
	\frac{\sqrt{z^2-x_{{\rm rec}}^{\; 2}}}{{\rm exp}(z)+1}
	\; ,
	\end{eqnarray}
	where $x_{{\rm rec}}=m_\chi^{\prime}\,(4/11\, \varepsilon\vert_{\rm rec})^{-1/3}/T_{\rm rec}$
	at recombination. 
	The impact on $\Delta N_{\rm eff}$ is
	maximized for $m_{\chi^{\prime}} \sim T_{\rm rec}$. 
	For example, if $m_{\chi^{\prime}}=0.6~$eV then $N_{\rm eff}=3.14$, which is in
	agreement with the value obtained by Planck \cite{Ade:2015xua}. 
	We speculate that values in this mass range 
	may also explain the tension concerning the decrease of $\Delta N_\mathrm{eff}$
	from observations related to big bang nucleosynthesis to observations of the 
	cosmic microwave background radiation. The dark sector suggested in this work 
	can explain this difference should it become a manifest tension. 
	\end{itemize}

\begin{table*}
  \begin{center}
    \resizebox{\textwidth}{!}{\begin{tabular}{r|c|c|c|c|c|c|c|}
      \cline{2-8}
      \multirow{2}{*}{}&\multirow{2}{*}{{$N=1$}}&\multicolumn{6}{|c|}{{$N=2$}}\\ \cline{3-8}
      &&I&II&III&IV&V&VI\\
  \cline{2-8}
  \multicolumn{1}{l|}{\begin{tikzpicture}[remember picture, baseline=-5pt]
   \node at (0,0)  (n1) {};
  \end{tikzpicture}}&\multirow{2}{*}{$m_{\chi}^{\left(1\right)}$}&\multirow{2}{*}{$m_{\chi^{\prime}}^{\left(1\right)},m_{\chi}^{\left(2\right)},m_{\chi}^{\left(1\right)}$}&\multirow{2}{*}{$m_{\chi}^{\left(1\right)},m_{\chi}^{\left(2\right)}$}&\multirow{2}{*}{$m_{\chi}^{\left(1\right)}$}&\multirow{2}{*}{$m_{\chi}^{\left(1\right)},m_{\chi}^{\left(2\right)}$}&\multirow{2}{*}{$m_{\chi}^{\left(1\right)}$}&\multirow{2}{*}{$m_{\chi}^{\left(1\right)}$}\\ 
      \multicolumn{1}{r|}{\multirow{2}{*}{{$T_{\rm{}D}$}}}&&&&&&&\\ \cline{2-8}
  
  \multicolumn{1}{r|}{}&\multirow{2}{*}{}&\multirow{2}{*}{}&\multirow{2}{*}{$m_{\chi^{\prime}}^{\left(1\right)}$}&\multirow{2}{*}{$m_{\chi^{\prime}}^{\left(1\right)},m_{\chi}^{\left(2\right)}$}&\multirow{2}{*}{}&\multirow{2}{*}{}&\multirow{2}{*}{$m_{\chi}^{\left(2\right)}$}\\ 
      \multicolumn{1}{r|}{\multirow{2}{*}{{$T_{\rm{}BBN}$}}}&&&&&&&\\ \cline{2-8}
  
      \multicolumn{1}{l|}{\begin{tikzpicture}[remember picture,baseline=5pt]
  \node (n2) {};
  \end{tikzpicture}}&\multirow{2}{*}{$m_{\chi^{\prime}}^{\left(1\right)}$}&\multirow{2}{*}{$m_{\chi^{\prime}}^{\left(2\right)}$}&\multirow{2}{*}{$m_{\chi^{\prime}}^{\left(2\right)}$}&\multirow{2}{*}{$m_{\chi^{\prime}}^{\left(2\right)}$}&\multirow{2}{*}{$m_{\chi^{\prime}}^{\left(1\right)},m_{\chi^{\prime}}^{\left(2\right)}$}&\multirow{2}{*}{$m_{\chi^{\prime}}^{\left(1\right)},m_{\chi}^{\left(2\right)},m_{\chi^{\prime}}^{\left(2\right)}$}&\multirow{2}{*}{$m_{\chi^{\prime}}^{\left(1\right)},m_{\chi^{\prime}}^{\left(2\right)}$}\\
      \multicolumn{1}{r|}{}&&&&&&&\\ \cline{2-8} \hline
  \multicolumn{1}{|r|}{$\left.\Delta{}N_{\rm{}eff}\right|_{\rm{}BBN}$}&0.21&0.21&0.38&0.57&0.30&0.39&0.45\\
             \hline
\multicolumn{1}{|r|}{$\left.\Delta{}N_{\rm{}eff}\right|_{\rm{}BBN}$}&0.17&0.17&0.31&0.46&0.27&0.36&0.40\\
             \hline\end{tabular}}
  \end{center}
  \caption{Mass spectra of dark fermions in accordance with our assumptions. 
    The first row of values for
    {$\left.\Delta{}N_{\rm{}eff}\right|_{\rm{}BBN}$}
     is for the case
    where the massive dark mediator is non-relativistic at the time of
    big bang nucleosynthesis, the second row for the case where it is relativistic.}\label{mass-spectra}
  \begin{tikzpicture}[remember picture, overlay]
  \draw[->,very thick] (n2) to node [sloped, above] {{$T$}} (n1);
  \end{tikzpicture}
  \end{table*}

\section{Cosmological observables\label{cosmological-observables}}
 
In this section we discuss the most important cosmological features
for a dark matter particle candidate: (A) the relic abundance, (B) the virtue of self-interactions, and
(C) the characteristic damping scales. 
In particular we comment on observational consequences of these features 
and give optimal values for the corresponding observables in relation to the parameters 
of the dark sector (\ref{ds}). An explicit benchmark point in parameter space is then 
presented in Section \ref{results}.

\subsection{The relic abundance}\label{chemical-decoupling}
A condition on the couplings of the dark sector can be found by requiring
that the dark matter particle candidates have the correct relic density. 
By assumption (\ref{assum3}) the relic density is dominated by the 
heavy fermions. 
For $T<m_{\chi}^{\left(i\right)}$ the dark matter particles annihilate 
rapidly through the dominant $s$-wave channels of the dark sector,
provided the hierarchy in the considered mass spectrum allows it, 
see Table \ref{mass-spectra}. The annihilation rate 
is approximately given by

\begin{eqnarray}
  \tfrac{m_{\chi}^{\left(i\right)\,{}2}}{\pi}\left\langle{}v_{\rm{}rel}\sigma_{\rm{}ann}\right\rangle_{i}
  &\approx{}&\
  \Theta\left(m_{\chi}^{\left(i\right)}-M_{X}\right)\alpha_{ii}^{2}
  \nonumber\\
  &&+\sum_{j=1}^{N}\Theta\left(m_{\chi}^{\left(i\right)}-m_{\chi}^{\left(j\right)}\right)\alpha_{ij}^{2}+\sum_{j^\prime=1}^{N}
  \alpha_{ij^{\prime}}^{2}\;{},
  \label{ann-cross-section}
\end{eqnarray} 
where {$\langle ... \rangle$} denotes the thermal average.  The first
term in Eq. (\ref{ann-cross-section}) is due to annihilations into
dark abelian gauge bosons, the second term is due to co-annihilations
among the heavier fermions (only permitted downwards the mass
hierarchy) and the third one is due to annihilations into the light
dark fermions (which is always downwards the mass hierarchy).  For
simplicity we assumed $M_{\tilde{h}}>m_{\chi}^{\left(i\right)}$, since
annihilations in the dark scalar field are $p$-wave suppressed.  The
annihilation rate (\ref{ann-cross-section}) is only a first order
approximation and admits corrections of order {$(Q_{\tilde{h}}\,
  m^{(i)}_\chi{}/v_{\rm{d}})^{2}$}.  These corrections are subdominant
for $m_{\chi}^{\left(i\right)} \lesssim v_{\rm{d}}$.  In this work we
consider observables only up to $Q_{\tilde{h}}^2$ corrections, which
is why we neglect annihilations into Standard Model particles after
the electroweak phase transition.

In order to obtain an expression for the relic density, we
start from the Boltzmann equation for the phase-space distribution function of dark matter and
solve it numerically following \cite{Balducci:2017vwg}. We find 
\begin{eqnarray}
	\label{om}
	\Omega_{i} h^2 &\approx& 0.06\; \left(\frac{0.13}{\alpha_i(
          N)}\right)^{2}\;
        \left(\frac{m_{\chi}^{\left(i\right)} }{{10\, \rm TeV}} \right)^2 \; ,
\end{eqnarray}	
where 
\begin{eqnarray}
	\alpha_i\left(N\right)
	&=&
	\left[\Theta (m_{\chi}^{\left(i\right)}-M_X)\,\alpha^2_{ii}
	+\sum_{j,j^\prime=1}^{N}
	\left(
	\Theta (m_{\chi}^{\left(i\right)}-m_{\chi}^{(j)})\,\alpha_{ij}^{2}
    + \alpha_{ij^{\prime}}^{2}\right)\right]^{1/2}\,{}.
\end{eqnarray}
This result includes the Sommerfeld effect
\cite{vandenAarssen:2012ag}, which gives order one corrections
towards smaller couplings. Note that the
self-annihilation cross sections are well below the experimental
sensitivity \cite{Abbasi:2012ws}.

\subsection{Self-interactions}\label{self-interactions}
 
The dark sector (\ref{ds})
includes self-interaction processes. Appropriate
values of the corresponding cross sections have been shown to solve
the so-called cusp versus core and too big to fail problems 
\cite{Vogelsberger:2012ku,Zavala:2012us,Tulin:2017ara}. Furthermore,
they also alleviate the so-called diversity problem
\cite{Kaplinghat:2015aga}. Indeed different formation histories of
galaxies, and thus different values for the baryon density {$\rho_{b}$}, lead
to different solutions for the isothermal density profile in the inner
halo region, which is obtained from the equation
{$\sigma_{0}^{2}\nabla^{2}\ln\left(\rho\right)=-4\pi{}G\left(\rho{}+\rho_{b}\right)$},
where {$\sigma_{0}$} is the one-dimensional velocity dispersion
\cite{Kaplinghat:2015aga}. The required values for the self-interaction cross sections are
$\langle\sigma_T/m_\ell^i\rangle_{v_{\rm therm}}\sim
1\,{}$cm$^2$g$^{-1}$ at the scale of dwarf galaxies and
$\langle\sigma_T/m_\ell^i\rangle_{v_{\rm therm}}\sim
0.1\,{}$cm$^2$g$^{-1}$ at the scale of clusters
\cite{Tulin:2017ara}. In these expressions the cross sections are
averaged over a thermal Maxwell-Boltzmann distribution with
{$v_{\rm{therm}}$} denoting the most probable velocity.  

The dark sector (\ref{ds}) includes self-interactions between the dark matter 
particles and is an example for a self-interacting dark matter theory. 
The averaged cross-section for self-interaction
per unit mass, $\left\langle{}\sigma_T/m^i_\ell\right\rangle_{v_{\rm therm}}$, is strongly velocity-dependent.
Numerical solutions for the momentum transfer between non-relativistic fermions can be found in \cite{Tulin:2013teo,Feng:2010zp}.
We make good use of the {\tt ETHOS} results  \cite{Cyr-Racine:2015ihg}. 
Note that the dark sector (\ref{ds}) combines self-interaction virtues with a late kinetic decoupling of
the dark matter particles which helps to address
the cusp versus core and too big to fail problems \cite{Schewtschenko:2015rno,Vogelsberger:2015gpr}.

\subsection{Damping masses}

The heavier fermions of the dark sector remain in kinetic equilibrium
with the lighter ones longer than with the gauge bosons if one assumes
$M_X \gg m_{\chi^{\prime}}^{\left(j\right)}$ for all
{$j\in\left\{1,\ldots{},N\right\}$}. At temperatures $M_X\gg T \gg
\text{max}_{j\in\left\{1,\ldots{},N\right\}}m_{\chi^{\prime}}^{\left(j\right)}$
and at lowest order in perturbation theory, the averaged
momentum-transfer elastic cross section per species is

 \begin{eqnarray}
	  \langle v_{\rm rel} \sigma_T \rangle_{\chi^{\left(i\right)}\chi^{\prime\,{}\left(j\right)}\rightarrow{}\chi^{\left(i\right)}\chi^{\prime\,{}\left(j\right)}} \approx \frac{40 \zeta(5) }{
            \pi \zeta(3)} G_{ij^{\prime}}^2 T^2 \left(\frac{T_{\rm
              DS}}{T}\right)^2 \;{}.
\end{eqnarray}
$T$ is the average photon temperature and $T_{\rm DS}$ the dark sector
 temperature. $\sigma_T$ is defined as $\sigma_{ T}\equiv \int
 \mathrm{d}\Omega (1-\cos \theta) \mathrm{d}\sigma_{\rm
   el}/\mathrm{d}\Omega$ and $G_{ij^{\prime}} \equiv
   \sqrt{2}g_ig_j^{\prime}/4M_X^2$ is the Fermi constant associated
 to the corresponding interaction. In the above expression
 {$m_{\chi^{\prime}}^{\left(j\right)}$} was considered
   negligible when compared to the other masses involved.

Before kinetic decoupling, the effective elastic interactions damp
perturbations in the linear power spectrum, which would otherwise grow
and form the first DM structures, i.e. protohalos
\cite{Shoemaker:2013tda}. For {$T<T_{\rm{kd}}$}, where {$T_{\rm{}kd}$}
is the temperature of kinetic decoupling, the interactions are too
weak to keep sustaining local thermal equilibrium and the remaining
elastic scatterings can be described as sources of entropy in an
imperfect DM fluid \cite{Hofmann:2001bi}. The corresponding damping
masses can be estimated by

\begin{equation}
M_{\rm d}=(4\pi/3) \rho_{\rm m}(T_{\rm{}kd})/H^3(T_{\rm kd})\;{},\label{damping-mass}
\end{equation}
since all the DM candidates within the Hubble radius at kinetic
decoupling were in thermal contact with the dark radiation plasma. In
general, collisionless damping or free streaming should be taken into
account as well. However, they are negligible for the values of
{$m_{\chi}^{\left(i\right)}$} considered here (see assumption
(\ref{assum3})). As stated in
Refs. \cite{Vogelsberger:2015gpr,Boehm:2000gq,Aarssen:2012fx}, the
missing satellite problem can be solved by damping masses of the order
of dwarf galaxies, i.e. of approximately {$10^{8-9}M_{\odot}$}. Thus,
by Eq. (\ref{damping-mass}), a temperature of kinetic decoupling in
the keV range is required
\cite{Tulin:2017ara,Karananas:2018goc,Hager:2020une}. Such values are
not excluded by any current collider and DM direct-search constraints
and can be accomodated very well in this model.  Furthermore, a
kinetic decoupling happening after BBN but before the sub-keV epoch
satisfies all Lyman-{$\alpha$} constraints
\cite{Aarssen:2012fx,Shoemaker:2013tda,Bullock:2010uy,Baur:2015jsy,
  Balducci:2017vwg}. More recent constraints are presented in Ref.
\cite{Irsic:2017ixq}, but they might be overly restrictive. We point
out that {$\Lambda{}$}CDM theories usually obtain damping masses in
the range of the earth mass \cite{Green:2005fa}, which lies well below
the resolution capabilities of current numerical simulations
\cite{Gondolo:2016mrz}.  An analytic derivation of the expression for
the temperature of kinetic decoupling {$T_{\rm{kd}}$} given a specific
particle model can be found in Ref.  \cite{Bringmann:2006mu}. Applying
this prescription to the case $N=1$ or equivalently for any $N$ as
long as the relic abundance is fixed through the freeze out of the
{$i$}-th dark lepton, we obtain
\begin{equation} \label{kinetic decoupling}
T_{\rm kd} \approx 147 \, {\rm eV} \, \varepsilon\vert_{T_{\rm kd}
}^{-1/2} \left(\frac{{10\, \rm TeV}}{m_{\chi}^{\left(i\right)}}\right)^{\frac{1}{4}}
\left(\frac{\Omega_{i}h^2}{{0.12}}\right)^{1/4}
\left(\frac{Q_{\tilde{h}}}{2\times 10^{-8}}\right) \left(\frac{v_{\rm d}}{{40\, \rm TeV}}\right)
\left(\frac{T}{T_\nu}\right)^{3/2}\;{}.
\end{equation}
Note that this result does not depend directly on {$\alpha(N)$}, which
is fixed by (\ref{om}).  Changes in {$\Omega_{i}$} do not play a big
role either. The strongest dependece of {$T_{\rm{kd}}$} is the one
with respect to {$Q_{\tilde{h}}$} and {$v_{\rm{}d}$}.

\section{Results and conclusion}\label{results}
\subsection{Results}
We now present a choice of parameters of the theory solving the cusp
vs core, too big to fail, missing satellites, and diversity problems
of {$\Lambda$}CDM. As a benchmark point we can choose for $N=1$
thermally produced WIMPs with a mass
{$m_{\chi}^{\left(1\right)}=10\,{}$}TeV and a mediator mass of
{$M_{X}=1.1\,{}$}MeV. This corresponds to a kinetic decoupling
temperature of {$T_{\rm{kd}}\approx{}0.456\,{}$}keV, yielding the
correct damping masses. The related cross sections for self
interactions are then in the desired range given in Section
\ref{self-interactions}. Given the number of conditions to be
satisfied (correct relic density, temperature of kinetic decoupling
and cross section for self interactions) the choice of parameters is
tightly constrained for one flavour of dark matter candidates. The
case of $N=2$ appears more flexible (for example following the
spectrum hierarchy proposed in section \ref{constraints}), since two
different generation couplings are present and even masses around 2
TeV seem to solve the small- and large-scale problems simultaneously.

\subsection{Conclusion}
In this work we studied the footprint of multiple flavours of
fermionic locally charged dark matter on the formation of small-scale
cosmic structures. This theory could be interpreted as the low-energy
part of the dark sector. In particular we showed that the underlying
theory satisfies all cosmological and particle-physics constraints and
solves the usual issues of structure formation: the cusp vs core, too
big to fail, missing satellites, and diversity problems. In order to
construct the theory, we introduced a SM-singlet family of flavored DM
consisting of light and heavy fields and we found that one should
introduce a new energy scale of interactions below the weak one in
order to obtain the required cosmological signatures. It should
however be clear that even larger groups as for example SU(2) could
straightforwardly lead to similar results and, therefore, we only
considered this U{$\left(1\right)$} theory as a starting point.  We
consider the model as very promising, since it can comfortably
accomodate all the important constraints.

\begin{acknowledgments}
   We appreciate financial support of our work by the DFG Excellence
   cluster ORIGINS.
\end{acknowledgments}

\bibliographystyle{unsrt}
%%\bibliography{draft}

\end{document}